# Collusion attack and counterattack on the quantum key agreement via non-maximally entangled cluster states

Jun Gu[1], Tzonelih Hwang[*]

*Department of Computer Science and Information Engineering, National Cheng Kung University, No. 1, University Rd., Tainan City, 70101, Taiwan, R.O.C.*

[1] isgujun@163.com

[*]**Corresponding Author:**

Tzonelih Hwang

Distinguished Professor

Department of Computer Science and Information Engineering,

National Cheng Kung University,

No. 1, University Rd.,

Tainan City, 70101, Taiwan, R.O.C.

Email: hwangtl@csie.ncku.edu.tw

TEL: +886-6-2757575 ext. 62524

# Abstract

Recently, Li et al. (Int J Theor Phys: DOI: 10.1007/s10773-020-04588-w, 2020) proposed a multiparty quantum key agreement protocol via non-maximally entangled cluster states. They claimed that the proposed protocol can help all the involved participants have equal influence on the final shared key. However, this study points out a loophole that makes Li et al.'s protocol suffer from a collusion attack, i.e. several dishonest participants can conspire to manipulate the final shared key without being detected by others. To avoid this loophole, an improvement is proposed here.

**Keywords** Quantum cryptography. Multiparty quantum key agreement. Collusion attack.

# 1. Introduction

In 1984, Bennet et al. proposed the first quantum key distribution (QKD) protocol [1]. Afterward, several QKD protocols [2, 3] and related protocols [4, 5] have been proposed. In these QKD protocols, the final shared key is first determined by a participant or a third-party and then distributed to the others. Different from the QKD, to ensure that all the participants have equal contribution to a final shared key, the quantum key agreement [6] (QKA) was proposed. In the QKA, none proper subset of the involved participants can determine any part of the final shared key without being detected by others.

Recently, Li et al. [7] proposed a multiparty quantum key agreement (MQKA) protocol via non-maximally entangled cluster states. They claimed that, in their MQKA protocol, all the involved participants have equal influences on the final shared key. However, this study shows that Li et al.'s MQKA protocol cannot avoid the collusion attack, i.e. two or three participants can determine the final secret key by themselves without being detected.

The rest of this paper is organized as follows. In Section 2, Li et al.'s MQKA protocol is reviewed. In Section 3, we show that Li et al.'s MQKA protocol suffers from a collusion attack and then propose a modified method to solve this problem. At last, a conclusion is given in Section 4.

## 2. A brief review of Li et al.'s MQKA

Before reviewing Li et al.'s MQKA protocol [7], some background is introduced first here.

**2.1 Background**

Four unitary operations $\{I,\sigma_z,\sigma_x,i\sigma_y\}$ and sixteen non-maximally entangled cluster states $\{\,|\Psi_i\rangle,\ (1\leq i\leq16)\,\}$ are used in Li et al.'s MQKA protocol. The $\{I,\sigma_z,\sigma_x,i\sigma_y\}$ and $|\Psi_i\rangle$ can be presented as follows:

$$\begin{cases} I = |0\rangle\langle 0|+|1\rangle\langle 1| = \begin{bmatrix} 1 & 0 \\ 0 & 1 \end{bmatrix} \\ \sigma_z = |0\rangle\langle 0|-|1\rangle\langle 1| = \begin{bmatrix} 1 & 0 \\ 0 & -1 \end{bmatrix} \\ \sigma_x = |1\rangle\langle 0|+|0\rangle\langle 1| = \begin{bmatrix} 0 & 1 \\ 1 & 0 \end{bmatrix} \\ i\sigma_y = |0\rangle\langle 1|-|1\rangle\langle 0| = \begin{bmatrix} 0 & 1 \\ -1 & 0 \end{bmatrix} \end{cases} \quad (1)$$

$$|\Psi_1\rangle_{1234} = (a|0000\rangle + b|0011\rangle + c|1100\rangle - d|1111\rangle)_{1234}$$

$$|\Psi_2\rangle_{1234} = (a|0000\rangle - b|0011\rangle + c|1100\rangle + d|1111\rangle)_{1234}$$

$$|\Psi_3\rangle_{1234} = (a|0000\rangle + b|0011\rangle - c|1100\rangle + d|1111\rangle)_{1234}$$

$$|\Psi_4\rangle_{1234} = (a|0000\rangle - b|0011\rangle - c|1100\rangle - d|1111\rangle)_{1234}$$

$$|\Psi_5\rangle_{1234} = (a|0001\rangle + b|0010\rangle - c|1101\rangle + d|1110\rangle)_{1234}$$

$$|\Psi_6\rangle_{1234} = (a|0001\rangle - b|0010\rangle - c|1101\rangle - d|1110\rangle)_{1234}$$

$$|\Psi_7\rangle_{1234} = (a|0001\rangle + b|0010\rangle + c|1101\rangle - d|1110\rangle)_{1234}$$

$$|\Psi_8\rangle_{1234} = (a|0001\rangle - b|0010\rangle + c|1101\rangle + d|1110\rangle)_{1234}$$

$$|\Psi_9\rangle_{1234} = (a|0100\rangle - b|0111\rangle + c|1000\rangle + d|1011\rangle)_{1234} \quad (2)$$

$$|\Psi_{10}\rangle_{1234} = (a|0100\rangle + b|0111\rangle + c|1000\rangle - d|1011\rangle)_{1234}$$

$$|\Psi_{11}\rangle_{1234} = (a|0100\rangle - b|0111\rangle - c|1000\rangle - d|1011\rangle)_{1234}$$

$$|\Psi_{12}\rangle_{1234} = (a|0100\rangle + b|0111\rangle - c|1000\rangle + d|1011\rangle)_{1234}$$

$$|\Psi_{13}\rangle_{1234} = (a|0101\rangle - b|0110\rangle - c|1001\rangle - d|1010\rangle)_{1234}$$

$$|\Psi_{14}\rangle_{1234} = (a|0101\rangle + b|0110\rangle - c|1001\rangle + d|1010\rangle)_{1234}$$

$$|\Psi_{15}\rangle_{1234} = (a|0101\rangle - b|0110\rangle + c|1001\rangle + d|1010\rangle)_{1234}$$

$$|\Psi_{16}\rangle_{1234} = (a|0101\rangle + b|0110\rangle + c|1001\rangle - d|1010\rangle)_{1234}$$

Each non-maximally entangled cluster state can be transformed into another one by performing the unitary operations $\{I, \sigma_z, \sigma_x, i\sigma_y\}$ on the 2nd and 4th particles. In Li et al.'s MQKA protocol, the encoding rule can be described as shown in Table 1.

Table 1. Encoding rule

| Unitary operations | Key | Unitary operations | Key |
| --- | --- | --- | --- |
| $I_2 \otimes I_4$ | 0000 | $(\sigma_x)_2 \otimes (\sigma_z)_4$ | 1000 |
| $I_2 \otimes (\sigma_z)_4$ | 0001 | $(\sigma_x)_2 \otimes I_4$ | 1001 |
| $(\sigma_z)_2 \otimes I_4$ | 0010 | $(i\sigma_y)_2 \otimes (\sigma_z)_4$ | 1010 |
| $(\sigma_z)_2 \otimes (\sigma_z)_4$ | 0011 | $(i\sigma_y)_2 \otimes I_4$ | 1011 |
| $(\sigma_z)_2 \otimes (\sigma_x)_4$ | 0100 | $(i\sigma_y)_2 \otimes (i\sigma_y)_4$ | 1100 |
| $(\sigma_z)_2 \otimes (i\sigma_y)_4$ | 0101 | $(i\sigma_y)_2 \otimes (\sigma_x)_4$ | 1101 |
| $I_2 \otimes (\sigma_x)_4$ | 0110 | $(\sigma_x)_2 \otimes (i\sigma_y)_4$ | 1110 |
| $I_2 \otimes (i\sigma_y)_4$ | 0111 | $(\sigma_x)_2 \otimes (\sigma_x)_4$ | 1111 |

**2.2 Li et al.'s MQKA protocol**

Suppose that there are $N$ participants $P_i (1 \leq i \leq N)$ in Li et al.'s MQKA protocol and each participant has a $4n$-bit private key $k_i$. The protocol can be described step by step as follows.

**Step 1** $P_i (1 \leq i \leq N)$ prepares $n$ non-maximally entangled cluster states $C_{1234}^{i,j} = \{(p_1^{i,j}, p_2^{i,j}, p_3^{i,j}, p_4^{i,j}), 1 \leq j \leq n\}$ in $|\Psi_1\rangle$ and then divides them into four ordered sequences $S_1^i = (p_1^{i,1}, p_1^{i,2}, \cdots, p_1^{i,n})$, $S_2^i = (p_2^{i,1}, p_2^{i,2}, \cdots, p_2^{i,n})$, $S_3^i = (p_3^{i,1}, p_3^{i,2}, \cdots, p_3^{i,n})$ and $S_4^i = (p_4^{i,1}, p_4^{i,2}, \cdots, p_4^{i,n})$. Subsequently, $P_i$ randomly generates enough decoy photons which are randomly in four states $\{|0\rangle, |1\rangle, |+\rangle, |-\rangle\}$ and inserts them into $S_2^i$ and $S_4^i$ to obtain two new sequences $S_2^{i\prime}$ and $S_4^{i\prime}$. Then, $P_i$ sends $S_2^{i\prime}$ and $S_4^{i\prime}$ to the next participant $P_{(i+1) \bmod N}$.

**Step 2** Upon receiving the acknowledgment of receipt from $P_{(i+1) \bmod N}$ through the authenticated channel, $P_i$ announces the positions and the corresponding bases of the decoy photons. Then $P_{(i+1) \bmod N}$ measures the decoy photons with the correct basis and announces the measurement results. Subsequently, $P_i$ checks whether the measurement results are equal to the initial states or not. If the error rate exceeds a predetermined value, this protocol will be aborted. Otherwise, $P_{(i+1) \bmod N}$ discards the decoy photons in $S_2^{i\prime}$ and $S_4^{i\prime}$ to obtain $S_2^i$ and $S_4^i$.

**Step 3** $P_{(i+1) \bmod N}$ encodes his/her private key $k_i$ on $S_2^i$ and $S_4^i$ according to the encoding rule shown in Table 1. Subsequently, similar to Step 1, $P_{(i+1) \bmod N}$ inserts decoy photons into $S_2^i$ and $S_4^i$. Then, he/she sends them to the next participant $P_{(i+2)\mod N}$.

**Step 4** Similar to Step 2 and Step 3, the participants $P_{(i+2)\mod N}$, $P_{(i+3)\mod N}$, $\cdots$, $P_{(i+N-1)\mod N}$ do the same eavesdropping detection process and encoding process one by one.

**Step 5** After $P_i$ receives $S_2^{i\prime}$ and $S_4^{i\prime}$ sent back from $P_{(i+N-1)\mod N}$, $P_i$ and $P_{(i+N-1)\mod N}$ use decoy photons to check whether there is an eavesdropper in the qubits transmission process or not. If there is no eavesdropper, $P_i$ obtains $S_2^i$ and $S_4^i$ and then performs POVM measurement [7] on $\{S_1^i, S_2^i, S_3^i, S_4^i\}$ to get the measurement result $M$. According to the encoding rule, $M = k_{(i+1)\mod N} \oplus k_{(i+2)\mod N} \oplus \cdots \oplus k_{(i+N-1)\mod N}$. Then $P_i$ can get the final key $K = k_i \oplus M$.

# 3. Collusion attack on Li et al.'s MQKA protocol and an improvement

Li et al. claimed that none proper subset of the participants can manipulate $K$ without being detected by others. However, this section shows that two or three participants can manipulate the final shared key by a collusion attack. That is, if the number of participants involved in Li et al.'s protocol is even, two participants $P_i$ and $P_{\left(\frac{N}{2}+i\right)\mod N}$ can cooperate to manipulate the key. Similarly, if odd participants are involved, three participants $P_i$, $P_{\left(\frac{N-1}{2}+i\right)\mod N}$ and $P_{\left(\frac{N+1}{2}+i\right)\mod N}$ can cooperate to manipulate the final key. Then, to solve this problem, a solution is proposed.

## 3.1 The loophole in Li et al.'s MQKA protocol

The collusion attack on Li et al.'s MQKA protocol can be described as follows. From Step 2 to Step 4, all the involved participants encode their private keys on the other participants' cluster states. However, the private keys are directly encoded on the particles. This means that if a participant gets his/her own generated particles during the encoding processes, he/she can directly measure these particles to obtain the XOR result of the keys encoded on them by the involved participants. Moreover, to obtain his/her own particles without being detected, the particles owner can simply cooperate with the other participant who is involved in the eavesdropper detection on these particles. That is, after a participant finishes the eavesdropper detection and discards all the decoy photons, he/she sends these particles back to their owner. Subsequently, the owner measures them and obtains the XOR result of the keys encoded on them. If the cooperating participants can get all the other participants' keys during the encoding processes, then they can perform a fake private key on the other participants' particles to manipulate the final shared key. Here, for ease of description, we first use a four-participant protocol as an example to present the collusion attack and then generalize it to all situations.

For example, assume that, there are four participants $P_1$, $P_2$, $P_3$ and $P_4$ involved in the protocol. $P_1$ and $P_3$ are the dishonest participants who want to manipulate the final key. In the encoding processes, from Step 2 to Step 4, $P_1$ $(P_2, P_3, P_4)$ performs unitary operators on { $S_2^4 S_4^4 \left( S_2^1 S_4^1, S_2^2 S_4^2, S_2^3 S_4^3 \right)$ , $S_2^3 S_4^3 \left( S_2^4 S_4^4, S_2^1 S_4^1, S_2^2 S_4^2 \right)$  $S_2^2 S_4^2 \left( S_2^3 S_4^3, S_2^4 S_4^4, S_2^1 S_4^1 \right)$ } according to his/her private key $k_1 (k_2, k_3, k_4)$. The quantum transfer processes in this part can be described as Fig 1.1 →Fig 1.2 →Fig 1.3 →Fig 1.4 →Fig 1.1. In Step 4, after $P_2 (P_4)$ encodes his/her private key $k_2 (k_4)$ on $P_1$'s $(P_3$'s$)$ particles $S_2^1 S_4^1 \left( S_2^3 S_4^3 \right)$, he/she sends $S_2^1 S_4^1 \left( S_2^3 S_4^3 \right)$ to $P_3 (P_1)$ (as shown in Fig 1.2). Upon receiving $S_2^1 S_4^1 \left( S_2^3 S_4^3 \right)$, $P_3 (P_1)$ sends $S_2^1 S_4^1 \left( S_2^3 S_4^3 \right)$ back to $P_1 (P_3)$. Subsequently, $P_1 (P_3)$ performs POVM measurement on $\{S_1^1, S_2^1, S_3^1, S_4^1\}$ ($\{S_1^3, S_2^3, S_3^3, S_4^3\}$) to get $k_2$ ($k_4$). After this, $P_1$ and $P_3$ can cooperate to obtain the final secret key $K$ by the formula $K=k_1 \oplus k_2 \oplus k_3 \oplus k_4$. If they want to use $K'$ to be the final key instead, after $P_3(P_1)$ receives $S_2^4 S_4^4$ ($S_2^2 S_4^2$) (as Fig 1.4 shows), $P_3(P_1)$ performs unitary operators on $S_2^4 S_4^4$ ($S_2^2 S_4^2$) according to the values of $k_1 \oplus k_2 \oplus k_4 \oplus K'$ ($k_2 \oplus k_3 \oplus k_4 \oplus K'$). Subsequently, in Step 5, $P_4(P_2)$ performs POVM measurement on $\{S_1^4, S_2^4, S_3^4, S_4^4\}$ ($\{S_1^2, S_2^2, S_3^2, S_4^2\}$) to obtain the measurement result $M_4(M_2)$ and computes the final key by the formula $M_4 \oplus k_4$ ($M_2 \oplus k_2$). However, the calculation results of $M_4 \oplus k_4$ and $M_2 \oplus k_2$ by $P_4$ and $P_2$ respectively are equal to $K'$ (because of $M_4 = k_1 \oplus k_2 \oplus (k_1 \oplus k_2 \oplus k_4 \oplus K') = k_4 \oplus K'$ and $M_2 = k_3 \oplus k_4 \oplus (k_2 \oplus k_3 \oplus k_4 \oplus K') = k_2 \oplus K'$) which was decided by $P_1$ and $P_3$.

If an even number of participants are involved in Li et al.'s protocol, the collusion attack can be described as in Fig 2.1. That is, the participants $P_i$ and $P_{\left(\frac{N}{2}+i\right) \bmod N}$ can cooperate to manipulate the final key. If the number of participants is odd, the collusion attack can be described as in Fig 2.2. That is, three participants $P_i$, $P_{\left(\frac{N-1}{2}+i\right) \bmod N}$ and $P_{\left(\frac{N+1}{2}+i\right) \bmod N}$ can cooperate to manipulate the final key. (Here we can consider $P_{\left(\frac{N-1}{2}+i\right) \bmod N}$ and $P_{\left(\frac{N+1}{2}+i\right) \bmod N}$ as one participant.)

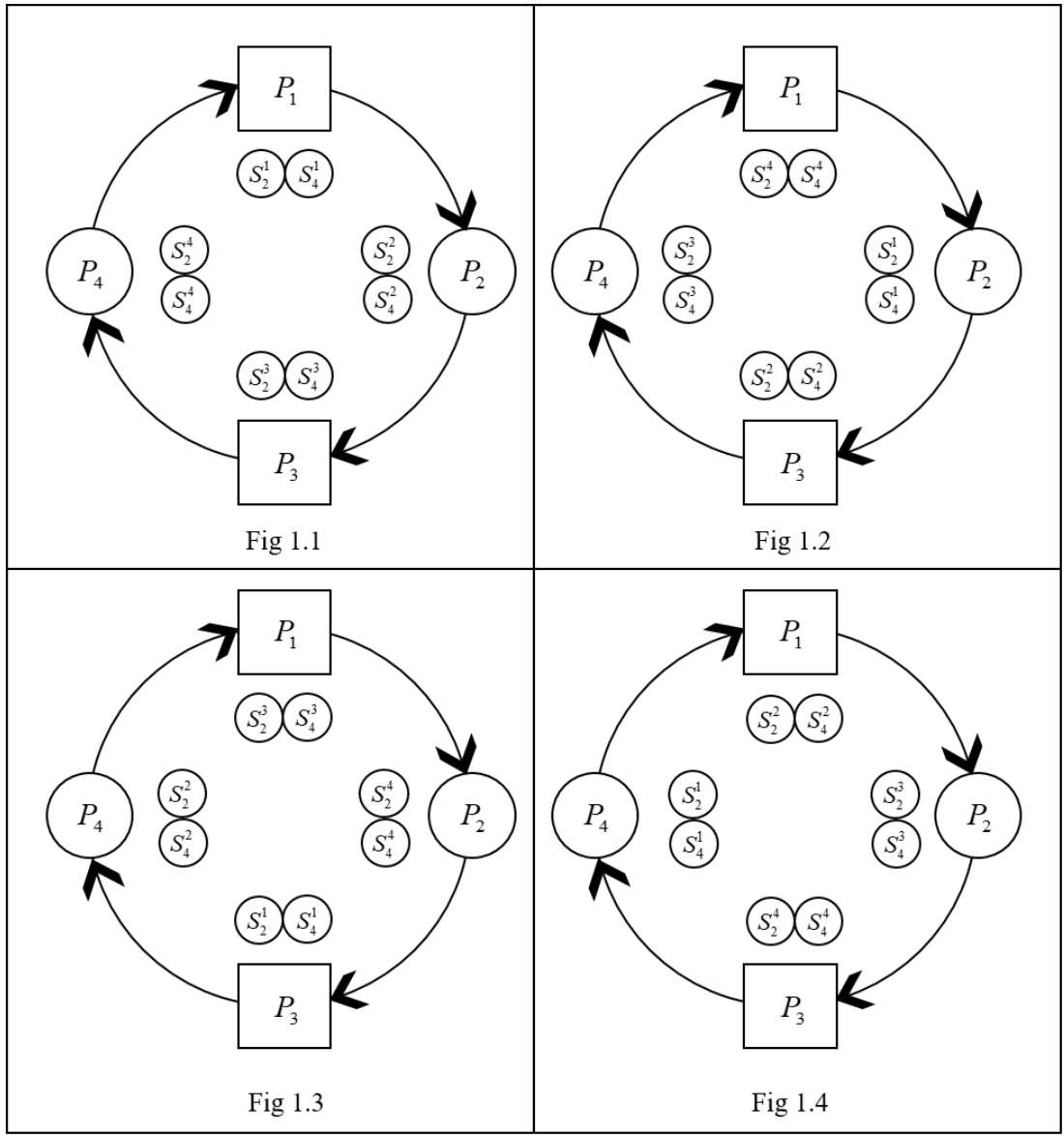

Fig 1. Collusion attack on Li et al.'s protocol of 4 participants

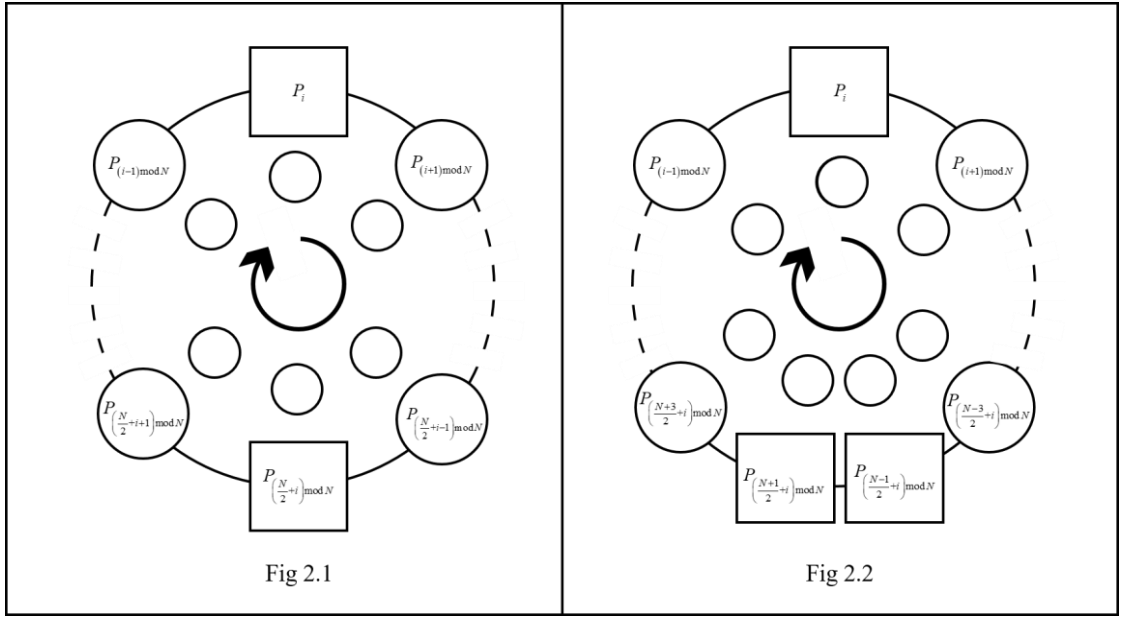

Fig 2. Collusion attack on Li et al.'s protocol

## 3.2 A solution to avoid the loophole

Because each participant's private key is directly encoded on the particles without protection, a dishonest participant can directly measure the particles generated by himself/herself to get the encoded key during the encoding processes. If all the participants randomly perform Hadamard operation $H = \frac{1}{\sqrt{2}}(|0\rangle\langle 0|+|0\rangle\langle 1|+|1\rangle\langle 0|-|1\rangle\langle 1|)$ or $I$ on each particle after encoding, without knowing the performed positions, these particles cannot be recovered to the initial bases. Hence, these particles cannot be directly measured by the particles owner anymore. However, for an entangled state, if a $H$ is performed before performing $\{\sigma_z, \sigma_x, i\sigma_y\}$, even if a $H$ is performed again, the result is still different from performing $\{\sigma_z, \sigma_x, i\sigma_y\}$ directly. Hence, to ensure that the Hadamard operation will not affect the correctness of encoding, we use four kinds of single photons $\{|0\rangle, |1\rangle, |+\rangle, |-\rangle\}$ as the quantum resource instead. Then, this loophole is solved. The detail of the improvement is as follows.

**Step 1\*** $P_i(1 \leq i \leq N)$ prepares a single photon sequence $S^i$ where every single photon is randomly chosen from $\{|0\rangle, |1\rangle, |+\rangle, |-\rangle\}$. Subsequently, similar to Step 1, $P_i$ generates enough decoy photons and inserts them into $S^i$ to obtain a new sequence $S^{i'}$. Then, $P_i$ sends $S^{i'}$ to the next participant $P_{(i+1) \mod N}$.

**Step 2\*** is the same as **Step 2** in Section 2.

**Step 3\*.** $P_{(i+1) \mod N}$ encodes his/her private key $k_i$ on $S^i$ by the unitary operations $I$ and $i\sigma_y$. That is, if the bit of $k_i$ is 1, $P_{(i+1) \mod N}$ performs $i\sigma_y$ on the particle. Otherwise, $P_{(i+1) \mod N}$ performs $I$ on it. Subsequently, for each particle, $P_{(i+1) \mod N}$ randomly performs $H$ or $I$ on it. Then, similar to Step 1\*, $P_{(i+1) \mod N}$ inserts decoy photons into $S^i$ and sends them to the next participant $P_{(i+2) \mod N}$.

**Step 4\*** is the same as **Step 4** in Section 2.

**Step 5\*.** After $P_i$ receives $S^{i'}$ sent back from $P_{(i+N-1) \mod N}$, $P_i$ and $P_{(i+N-1) \mod N}$ check whether there is an eavesdropper in the qubits transmission process or not. After all the participants finish the eavesdropper detection, all the participants announce the $H$ performed on $S_i$. Then, $P_i$ recovers the particles in $S_i$ to the initial bases and measures them. $P_i$ gets the measurement result $M$. According to the encoding rule, $M = k_{(i+1) \mod N} \oplus k_{(i+2) \mod N} \oplus \cdots \oplus k_{(i+N-1) \mod N}$. Then $P_i$ can compute the final key $K = k_i \oplus M$.

With this modified method, the collusion attack can be avoided.

## 4. Conclusions

Li et al. proposed a multiparty quantum key agreement protocol with non-maximally entangled cluster states. However, this study shows that Li et al.'s MQKA protocol suffers from a collusion attack. A solution with merely single photons is hence proposed here.

## Acknowledgment

We would like to thank the Ministry of Science and Technology of the Republic of China, Taiwan for partially supporting this research in finance under the Contract No. MOST 109-2221-E-006-168-; No. MOST 108-2221-E-006-107-.

## References

[1] Charles H. Bennet and Gilles Brassard, "Quantum cryptography: Public key distribution and coin tossing," in *Proceedings of the IEEE International Conference on Computers, Systems and Signal Processing, Bangalore, India*, 1984, pp. 175-179.

[2] Michel Boyer, Dan Kenigsberg, and Tal Mor, "Quantum key distribution with classical Bob," in *Quantum, Nano, and Micro Technologies, 2007. ICQNM'07. First International Conference on*, 2007, pp. 10-10: IEEE.

[3] Michel Boyer, Ran Gelles, Dan Kenigsberg, and Tal Mor, "Semiquantum key distribution," *Physical Review A,* vol. 79, no. 3, p. 032341, 2009.

[4] Hoi-Kwong Lo, Xiongfeng Ma, and Kai Chen, "Decoy state quantum key distribution," *Physical review letters,* vol. 94, no. 23, p. 230504, 2005.

[5] Renato Renner, "Security of quantum key distribution," *International Journal of Quantum Information,* vol. 6, no. 01, pp. 1-127, 2008.

[6] Nanrun Zhou, Guihua Zeng, and Jin Xiong, "Quantum key agreement protocol," *Electronics Letters,* vol. 40, no. 18, pp. 1149-1150, 2004.

[7] Taichao Li, Xu Wang, and Min Jiang, "Quantum Key Agreement Via Non-maximally Entangled Cluster States," *International Journal of Theoretical Physics,* pp. 1-16, 2020.